\documentclass[aps,preprint,showpacs]{revtex4}
\usepackage{amsmath}
\usepackage[english]{babel}
\usepackage[latin1]{inputenc}
\usepackage{amssymb}
\usepackage{graphicx}
\usepackage{natbib}
\usepackage{epsfig}
\usepackage{bm}
\usepackage{dcolumn}

\begin{document}

\vspace*{2cm}

\title{\sc\Large{Magnetic catalysis and inverse magnetic
catalysis in nonlocal chiral quark models}}

\author{V.P. Pagura$^{a}$, D. G\'omez Dumm$^{b,c}$, S. Noguera$^{a}$ and N.N.\ Scoccola$^{c,d,e}$}

\affiliation{$^{a}$ Departamento de F\'{\i}sica Te\'{o}rica and IFIC, Centro Mixto
Universidad de Valencia-CSIC, E-46100 Burjassot (Valencia), Spain}
\affiliation{$^{b}$ IFLP, CONICET $-$ Departamento de F\'{\i}sica, Fac.\ de Cs.\ Exactas,
Universidad Nacional de La Plata, C.C. 67, (1900) La Plata, Argentina}
\affiliation{$^{c}$ CONICET, Rivadavia 1917, (1033) Buenos Aires, Argentina}
\affiliation{$^{d}$ Physics Department, Comisi\'{o}n Nacional de Energ\'{\i}a At\'{o}mica, }
\affiliation{Av.\ Libertador 8250, (1429) Buenos Aires, Argentina}
\affiliation{$^{e}$ Universidad Favaloro, Sol{\'{\i}}s 453, (1078) Buenos Aires, Argentina}

\begin{abstract}
We study the behavior of strongly interacting matter under an external
constant magnetic field in the context of nonlocal chiral quark models
within the mean field approximation. We find that at zero temperature the
behavior of the quark condensates shows the expected magnetic catalysis
effect, our predictions being in good quantitative agreement with lattice
QCD results. On the other hand, in contrast to what happens in the standard
local Nambu$-$Jona-Lasinio model, when the analysis is extended to the case
of finite temperature our results show that nonlocal models naturally lead
to the Inverse Magnetic Catalysis effect.
\end{abstract}


\maketitle

\renewcommand{\thefootnote}{\arabic{footnote}}
\setcounter{footnote}{0}

Over the last few years, the understanding of the behavior of strongly
interacting matter under extremely intense magnetic fields has attracted
increasing attention, due to its relevance for various subjects such as the
physics of compact objects like magnetars~\cite{duncan}, the analysis of
heavy ion collisions at very high energies~\cite{HIC} and the study of the
first phases of the Universe~\cite{cosmo}. Consequently, considerable work
has been devoted to studying the structure of the QCD phase diagram in the
presence of an external magnetic field (see
Refs.~\cite{Kharzeev:2012ph,Andersen:2014xxa,Miransky:2015ava} for recent
reviews). On the basis of the results arising from most low-energy effective
models of QCD it was generally expected that, at zero chemical potential,
the magnetic field would lead to an enhancement of the chiral condensate
(``magnetic catalysis''), independently of the temperature of the system.
However, lattice QCD (LQCD) calculations carried out with physical pion
masses~\cite{Bali:2011qj,Bali:2012zg} show that, whereas at low temperatures
one finds indeed such an enhancement, the situation is quite different close
to the critical chiral restoration temperature: in that region light quark
condensates exhibit a nonmonotonic behavior as functions of the external
magnetic field, which results in a decrease of the transition temperature
when the magnetic field is increased. This effect is known as inverse
magnetic catalysis (IMC). Although many scenarios have been considered in
the last few years to account for the
IMC~\cite{Skokov:2011ib,Fraga:2012ev,Bruckmann:2013oba,Bali:2013esa,Fukushima:2012kc,
Chao:2013qpa,Fraga:2013ova,Ferreira:2013tba,Ferreira:2014kpa,Ayala:2014iba,Farias:2014eca,
Ayala:2014gwa,Fayazbakhsh:2014mca,Andersen:2014oaa,Mueller:2015fka,Ayala:2014uua,Ferrer:2014qka,
Braun:2014fua,Rougemont:2015oea,Ayala:2015bgv,Mao:2016fha}, the mechanism
behind this effect is not yet fully understood. With this motivation, in
this work we study the behavior of strongly interacting matter under an
external magnetic field in the framework of nonlocal chiral quark models.
These theories are proposed as a sort of nonlocal extensions of the
well-known Nambu$-$Jona-Lasinio (NJL) model, intending to go a step further
toward a more realistic effective approach to QCD. In fact, nonlocality
arises naturally in the context of successful descriptions of low-energy
quark dynamics~\cite{Schafer:1996wv,RW94}, and it has been
shown~\cite{Noguera:2008} that nonlocal models can lead to a momentum
dependence in quark propagators that is consistent with LQCD results.
Another advantage of these models is that the effective interaction is
finite to all orders in the loop expansion, and therefore there is no need
to introduce extra cutoffs~\cite{Rip97}. Moreover, in this framework it is
possible to obtain an adequate description of the properties of light mesons
at both zero and finite
temperature/density~\cite{Noguera:2008,Bowler:1994ir,Schmidt:1994di,Golli:1998rf,
General:2000zx,Scarpettini:2003fj,GomezDumm:2004sr,GomezDumm:2006vz,
Contrera:2007wu,Hell:2008cc,Dumm:2010hh}. A previous attempt of considering
the effect of an external magnetic field within these models was done in
Ref.~\cite{Kashiwa:2011js}. In that work the magnetic field was introduced
by using a simplified extension of the method usually followed in the local
NJL model, and no signal of IMC was found. In the present article we
concentrate on the analysis of nonlocal quark models with separable
interactions, including a coupling to a uniform magnetic field. We address
the problem by following a more rigorous procedure based on the Ritus
eigenfuncion method~\cite{Ritus:1978cj}, which allows us to properly obtain
the corresponding mean field action and to derive the associated gap
equation. Then we solve this equation numerically for different values of
the external magnetic field, considering the case of systems at both zero
and finite temperature. We find that at zero temperature the behavior of the
quark condensates shows the expected magnetic catalysis effect, our
predictions being in good quantitative agreement with LQCD results. On the
other hand, in contrast to what happens in the standard local
Nambu$-$Jona-Lasinio model, when the analysis is extended to the case of
finite temperature our results show that nonlocal models naturally lead to
the IMC effect already at the mean field level.

\hfill

\noindent{\em Theoretical formalism}

We begin by stating the Euclidean action for a simple nonlocal chiral quark
model that includes two light flavors,
\begin{equation}
S_E = \int d^4 x \ \left\{ \bar \psi (x) \left(- i \rlap/\partial
+ m_c \right) \psi (x) -
\frac{G}{2} j_a(x) j_a(x) \right\} \ .
\label{action}
\end{equation}
Here $m_c$ is the current quark mass, which is assumed to be equal
for $u$ and $d$ quarks. The nonlocal currents $j_a(x)$ are given by
\begin{eqnarray}
j_a (x) &=& \int d^4 z \  {\cal G}(z) \
\bar \psi(x+\frac{z}{2}) \ \Gamma_a \ \psi(x-\frac{z}{2}) \ ,
\label{cuOGE}
\end{eqnarray}
where $\Gamma_{a}=(\leavevmode\hbox{\small1\kern-3.8pt\normalsize1},i\gamma
_{5}\vec{\tau})$, and the function ${\cal G}(z)$ is a nonlocal form factor
that characterizes the effective interaction. Since we are interested in
studying the influence of a magnetic field, we introduce in the effective
action Eq.~(\ref{action}) a coupling to an external electromagnetic gauge
field $\mathcal{A}_{\mu}$. For a local theory this can be done by performing
the replacement $\partial_{\mu}\rightarrow\partial_{\mu}-i\ \hat Q
\mathcal{A}_{\mu}(x)$, where $\hat Q=\mbox{diag}(q_u,q_d)$, with $q_u=2e/3$,
$q_d = -e/3$, is the electromagnetic quark charge operator. In the case of
the nonlocal model under consideration the situation is more complicated
since the inclusion of gauge interactions implies a change not only in the
kinetic terms of the Lagrangian but also in the nonlocal currents in
Eq.~(\ref{cuOGE}). One has
\begin{equation}
\psi(x-z/2) \rightarrow W\left(  x,x-z/2\right)  \ \psi(x-z/2)\ ,
\end{equation}
and a related change holds for $\bar
\psi(x+z/2)$~\cite{GomezDumm:2006vz,Noguera:2008,Dumm:2010hh}. Here the
function $W(s,t)$ is defined by
\begin{equation}
W(s,t)\ =\ \mathrm{P}\;\exp\left[ -\, i \hat Q\int_{s}^{t}dr_{\mu}\  \mathcal{A}_{\mu
}(r)\right]  \ , \label{intpath}%
\end{equation}
where $r$ runs over an arbitrary path connecting $s$ with $t$. As is
usually done, we take it to be a straight line path.

To proceed we bosonize the fermionic theory, introducing scalar and
pseudoscalar fields $\sigma(x)$ and $\vec{\pi}(x)$ and integrating out the
fermion fields. The bosonized action can be written
as~\cite{Noguera:2008,Dumm:2010hh}
\begin{equation}
S_{\mathrm{bos}}=-\ln\det\mathcal{D}+\frac{1}{2G}
\int d^{4}x
\Big[\sigma(x)\sigma(x)+ \vec{\pi}(x)\cdot\vec{\pi}(x)\Big]
\ ,
\end{equation}
where
\begin{eqnarray}
\mathcal{D}\left(  x+\frac{z}{2}\,,x-\frac{z}{2}\right)   &  = &\gamma_{0}\;W\left(  x+\frac{z}{2},x\right)  \gamma_{0}
\, \bigg[\,\delta^{(4)}(z)\,\big(-i\rlap/\partial+m_{c}\big) + \nonumber \\
& & \qquad \mathcal{G}(z)\big[  \sigma\left(  x\right)+i\vec{\tau}\cdot\vec{\pi}\left(  x\right)  \big]
\bigg]\; W\left(  x,x-\frac{z}{2}\right) \ .
\label{aa}%
\end{eqnarray}
Let us consider the particular case of a constant and homogenous magnetic
field orientated along the 3-axis. Choosing the Landau gauge, the
corresponding gauge field is given by $\mathcal{A}_\mu = B\, x_1\,
\delta_{\mu 2}$. Next, we assume that the field $\sigma$ has a nontrivial
translational invariant mean field value $\bar{\sigma}$, while the mean
field values of pseudoscalar fields $\pi_{i}$ are zero. It should be
stressed at this point that the assumption that $\bar{\sigma}$ is
independent of $x$ does not imply that the resulting quark propagator will
be translational invariant. In fact, as discussed below, one can show that
such an invariance is broken by the appearance of the usual Schwinger phase.
Our assumption just states that the deviations from translational invariance
that are inherent to the magnetic field are not affected by the dynamics of
the theory. In this way, within the mean field approximation (MFA) we get
\begin{eqnarray}
\mathcal{D}^{\mbox{\tiny MFA}} (  x , x') &=& \delta^{(4)}(x-x') \left( - i \rlap/\partial
- \hat Q \, B \, x_1 \, \gamma_2 + m_c \right)  + \nonumber \\
& & \qquad
\bar \sigma \ \mathcal{G}(x-x') \; \exp\left[ \frac{i}{2} \, \hat Q \, B \, (x_2 - x_2')\, (x_1 +
x_1')\right]\ .
\end{eqnarray}

At this stage it is convenient to follow the Ritus eigenfuncion method
\cite{Ritus:1978cj}. Thus, we introduce the function
\begin{equation}
\mathcal{D}^{\mbox{\tiny MFA}}_{p,p'} = \int d^4x \ d^4x' \
\bar{\mathbb{E}}_{p} (x)  \ \mathcal{D}^{\mbox{\tiny MFA}}(x,x')  \
\mathbb{E}_{p'} (x')\ ,
\label{dpp}
\end{equation}
where $\mathbb{E}_{p}$ are the usual Ritus matrices, with
$p=(k,p_2,p_3,p_4)$. The r.h.s.\ of Eq.~(\ref{dpp}) can be worked out, and
after some calculation one arrives at a relatively compact expression
for $\mathcal{D}^{\mbox{\tiny MFA}}_{p,p'}$, which is shown to be diagonal
in flavor space. For each flavor $f=u,d$ one has
\begin{eqnarray}
\mathcal{D}^{\mbox{\tiny MFA},f}_{p,p'} & = & (2\pi)^4\, \delta_{kk'}\, \delta(p_2 - p_2\!')
\, \delta(p_3 - p_3\!') \, \delta(p_4 - p_4\!') \times \nonumber \\
& & \!\!\!\!\!\!\!\bigg[ \big[\mathbb{I} + \delta_{k0}(\Delta^{s_f} - \mathbb{I})\big]
\big(-s_f\sqrt{2 k\, |q_f B|}\; \gamma_2 + p_3\,\gamma_3 + p_4\,\gamma_4 \big) +
\sum_{\lambda=-1,1} \!\Delta^{\lambda} M^{\lambda,f}_{\bar p,k} \bigg] ,
\label{dmfa}
\end{eqnarray}
where we have defined $s_{\! f} = {\rm sign}(q_f B)$ and
$\Delta^{\lambda}=\mbox{diag}(\delta_{1\lambda},
\delta_{-1\lambda},\delta_{1\lambda},\delta_{-1\lambda})$, whereas
$M^{\lambda,f}_{\bar p,k}$ is given by
\begin{eqnarray}
M^{\lambda,f}_{\bar p,k} = (-1)^{k-\frac{1-\lambda\ \! \! s_{\! f}}{2}} \int_0^\infty dr
\, r \; \exp(-r^2/2) \left[ m_c + \bar \sigma \, g\bigg( \frac{|q_f B|}{2}
r^2 + \bar p^2\bigg)\right] \, L_{k-\frac{1-\lambda\ \! \!s_{\! f}}{2}}(r^2)\ .
\label{mpk}
\end{eqnarray}
Here $g(p^2)$ stands for the Fourier transform of ${\cal G}(z)$, $\bar
p=(p_3,p_4)$ is a two-dimensional vector and $L_n(x)$ are the Laguerre
polynomials. We use the standard convention $L_{-1}(x)=0$, hence
$M^{-s_{\! f},f}_{\bar p,0}=0$. From Eq.~(\ref{dmfa}) we finally find that the MFA
action per unit volume can be expressed as
\begin{eqnarray}
\frac{S^{\mbox{\tiny MFA}}_{\mathrm{bos}}}{V^{(4)}} & = & \frac{ \bar
\sigma^2}{2 G} - N_c \sum_{f=u,d} \frac{  |q_f B|}{2 \pi} \int \frac{d^2\bar
p}{(2\pi)^2} \ \Bigg\{ \ln\left[\bar p^2 + \left({M^{s_{\! f},f}_{\bar p,0}\,}\right)^2\right]
+ \nonumber \\
& & \qquad \sum_{k=1}^\infty \ \ln\left[ \left( 2 k |q_f B| + \bar p^2 +
M^{-1,f}_{\bar p,k} M^{+1,f}_{\bar p,k}\right)^2 \! \! + \bar p^2 \left(
M^{+1,f}_{\bar p,k} - M^{-1,f}_{\bar p,k} \right)^2\right]\Bigg\}\ .
\label{smfa}
\end{eqnarray}
The corresponding gap equation can be now easily found by minimizing this
expression with respect to $\bar \sigma$. It is worth mentioning that this
gap equation can be also obtained using the Schwinger-Dyson formalism for
the quark propagator discussed in e.g.\
Refs.~\cite{Leung:1996qy,Watson:2013ghq,Mueller:2014tea}. Actually, it turns
out that the two point function in Eq.~(\ref{dmfa}) can be casted into a
form similar to that given in Ref.~\cite{Watson:2013ghq}. Thus, using the
analysis discussed in that work, one can show that the associated quark
propagator in coordinate space can be written as the product of the
exponential of a Schwinger phase (which breaks translational invariance)
times a translational invariant function.

The above results can be now extended to finite temperature using the
Matsubara formalism. This amounts to performing the replacement
\begin{equation}
\int \frac{d^2\bar p}{(2\pi)^2}\; F(\bar p^2) \ \rightarrow \ T
\sum_{n=-\infty}^{\infty} \int \frac{dp_3}{2\pi}\; F(\bar p_n^{\,2})\ ,
\end{equation}
where $\bar p_n = (p_3,\omega_n)$, $\omega_n = (2n + 1) \pi T$ being the
Matsubara frequencies for fermionic modes. In this way one can easily obtain
the MFA finite temperature thermodynamical potential $\Omega^{\mbox{\tiny
MFA}}$, as well as the related gap equation. Given $\Omega^{\mbox{\tiny
MFA}}$, the magnetic field dependent quark condensate for each flavor can be
calculated by taking the derivative with respect to the corresponding
current quark mass. This leads to
\begin{eqnarray}
\langle \bar q_f q_f\rangle_{B,T} &=& - \frac{N_c\, |q_f B|\, T}{\pi}
\int \frac{d p_3}{2\pi} \sum_{k=0}^\infty \sum_{n=-\infty}^\infty
\nonumber \\
& & \qquad \frac{ M^{-1,f}_{\bar p_n,k} \left[\bar p_n^{\,2} + 2 k |q_f B|
+ {M^{+1,f}_{\bar p_n,k}}^2\right] \ + \ ( + \leftrightarrow - )}
{\left( 2 k |q_f B| + \bar p_n^{\,2} + M^{-1,f}_{\bar p_n,k}
M^{+1,f}_{\bar p_n,k}\right)^2 \! \! + \bar p_n^{\,2}
\left( M^{+1,f}_{\bar p_n,k} - M^{-1,f}_{\bar p_n,k} \right)^2}\ .
\label{condensate}
\end{eqnarray}

As it is usually found in the context of nonlocal
models~\cite{GomezDumm:2004sr}, this expression turns out to be divergent
beyond the chiral limit. We obtain a regularized condensate by subtracting
the corresponding expression in the absence of quark-quark interactions and
adding it in a regularized form. Thus we have
\begin{equation}
\langle \bar q_f q_f\rangle^{\rm reg}_{B,T} = \langle \bar q_f q_f\rangle_{B,T}
- \langle \bar q_f q_f\rangle^{\rm free}_{B,T} +
\langle \bar q_f q_f\rangle^{\rm free,reg}_{B,T}\ .
\end{equation}
Notice that ``free'' condensates are defined keeping the interaction with
the magnetic field. In the case of $\langle \bar q_f q_f\rangle^{\rm
free,reg}_{B,T}$ the Matsubara sum can be performed analytically, leading to
\begin{eqnarray}
\langle \bar q_f q_f\rangle^{\rm free,reg}_{B,T}
&=& - \frac{N_c m_c^3}{4 \pi^2}
\left[ \frac{\ln \Gamma(x_f)}{x_f} - \frac{\ln 2 \pi}{2 x_f} + 1 - \left(1-\frac{1}{2x_f}\right) \ln x_f \right]
+ \nonumber \\
& & \qquad \frac{N_c |q_f B|}{\pi} \sum_{k=0}^\infty \alpha_k
\int \frac{d p}{2\pi} \frac{m_c}{\epsilon^f_k \left[ 1 +
\exp(\epsilon^f_k/T)\right]}\ ,
\end{eqnarray}
where $x_f = m_c^2/(2 |q_f B|)$, $\alpha_k = 2-\delta_{k0}$ and
$\epsilon^f_k = \sqrt{ 2 k |q_f B| + p^2 + m_c^2}\,$. Finally, to make
contact with the LQCD results quoted in Ref.~\cite{Bali:2012zg} we define
the quantity
\begin{equation}
\Sigma^f_{B,T} = \frac{2\, m_c}{S^4} \left[ \langle \bar q_f q_f \rangle^{\rm reg}_{B,T}
 - \langle \bar q_f q_f\rangle^{\rm reg}_{0,0} \right] + 1 \ ,
\label{defi}
\end{equation}
where the scale $S$ is given by $S= (135 \times 86)^{1/2}$ MeV. We also
introduce the definitions $\Delta \Sigma^f_{B,T} =  \Sigma^f_{B,T} -
\Sigma^f_{0,T}$ and $\Delta \bar \Sigma_{B,T} = (\Delta
\Sigma^u_{B,T}+\Delta \Sigma^d_{B,T})/2\,$.

\hfill

\noindent {\em Numerical results}

\begin{figure}[hbt]
\includegraphics[width=0.5\textwidth]{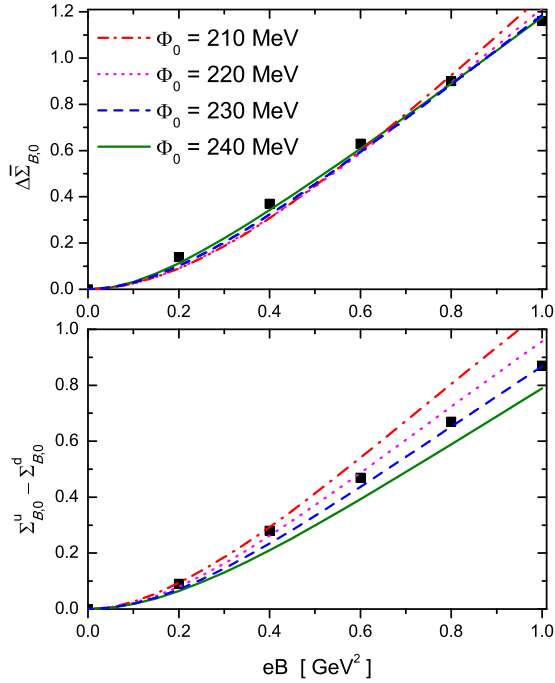}
\caption{Normalized condensates as functions of the magnetic field at $T =
0$. The curves correspond to different model parametrizations identified by
$\Phi_0 = (- \langle\bar q_f q_f \rangle^{\rm reg}_{0,0})^{1/3}$, for the
case of a Gaussian form factor. Full square symbols correspond to LQCD
results taken from Ref.~\cite{Bali:2012zg}. Upper panel: subtracted flavor
average; lower panel: flavor difference [see Eq.~(\ref{defi}) and the text
below].} \label{fig1}
\end{figure}

To obtain the numerical predictions that follow from the above formalism, it
is necessary to specify the particular form of the nonlocal form factor.
For simplicity, let us consider the often-used Gaussian form $g(p^2) =
\exp(-p^2/\Lambda^2)$, where the effective scale $\Lambda$ is an additional
parameter of the model. This form factor has the particular advantage that
the integral in Eq.~(\ref{mpk}) can be performed analytically. One gets
\begin{equation}
M^{\lambda,f}_{\bar p,k} = m_c + \bar \sigma \
\frac{ \left(1- |q_f B|/\Lambda^2\right)^{k+\frac{\lambda s_{\! f}-1}{2}}}
{ \left(1+ |q_f B|/\Lambda^2\right)^{k+\frac{\lambda s_{\! f}+1}{2}}}
\;\exp\!\big(-{\bar p}^{\,2}/\Lambda^2\big)\ .
\label{gauss}
\end{equation}
Our numerical results for $T=0$ are shown in Fig.~\ref{fig1}. In the upper
panel we quote the model predictions for $\Delta \bar \Sigma_{B,0}$ as a
function of $eB$ for various model parametrizations, while in the lower
panel we show the corresponding results for $\Sigma^u_{B,0}-\Sigma^d_{B,0}$.
LQCD data from Ref.~\cite{Bali:2012zg} are also displayed in both cases for
comparison. Note that the nonlocal model has three parameters, namely,
$m_c$, $G$ and $\Lambda$. They have been fixed to reproduce the empirical
values of the pion mass and decay constant, and to lead to a certain chosen
value of the quark condensate at zero temperature and magnetic field that we
identify by $\Phi_0 \equiv (- \langle\bar q_f q_f \rangle^{\rm
reg}_{0,0})^{1/3}$. Details of this parameter fixing procedure can be found
in Ref.~\cite{GomezDumm:2006vz}, where the explicit values of the parameters
for $\Phi_0 = 220$ MeV and $240$ MeV are given. As seen in Fig.~\ref{fig1},
the predictions for $\Delta \bar \Sigma_{B,0}$ are very similar for all
parametrizations considered, and show a very good agreement with LQCD
results. In the case of $\Sigma^u_{B,0}-\Sigma^d_{B,0}$ we see that,
although the overall agreement with LQCD calculations is still good, there
is a somewhat larger dependence on the model parametrization.

\begin{figure}[hbt]
\includegraphics[width=0.8\textwidth]{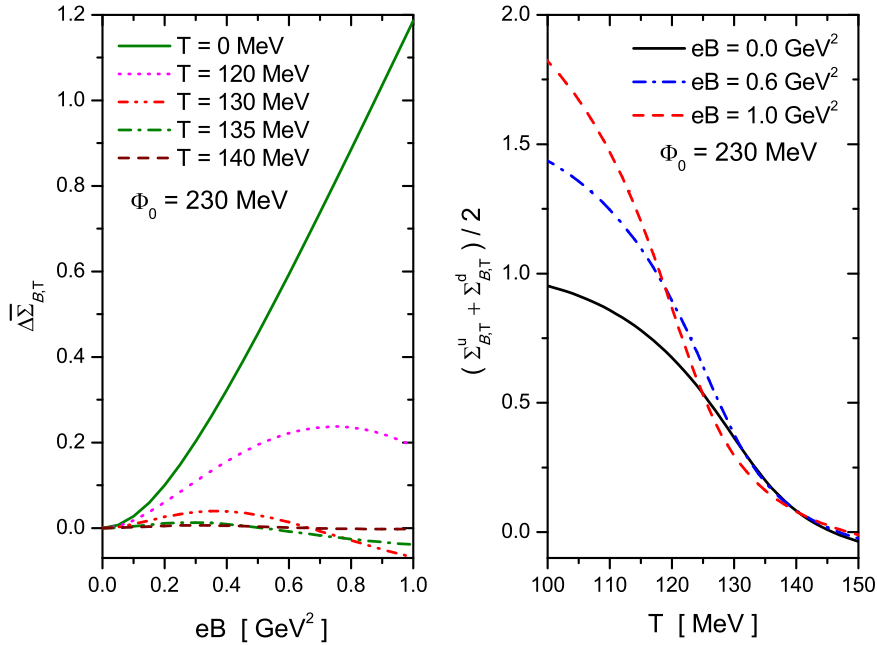}
\caption{Left: subtracted normalized flavor average condensate as a function
of $eB$ for different representative temperatures. Right: normalized flavor
average condensate as a function of the temperature for different
representative values of $eB$. Results in both panels correspond to
$\Phi_0=230$ MeV.}
\label{fig2}
\end{figure}

\begin{figure}[hbt]
\includegraphics[width=0.9\textwidth]{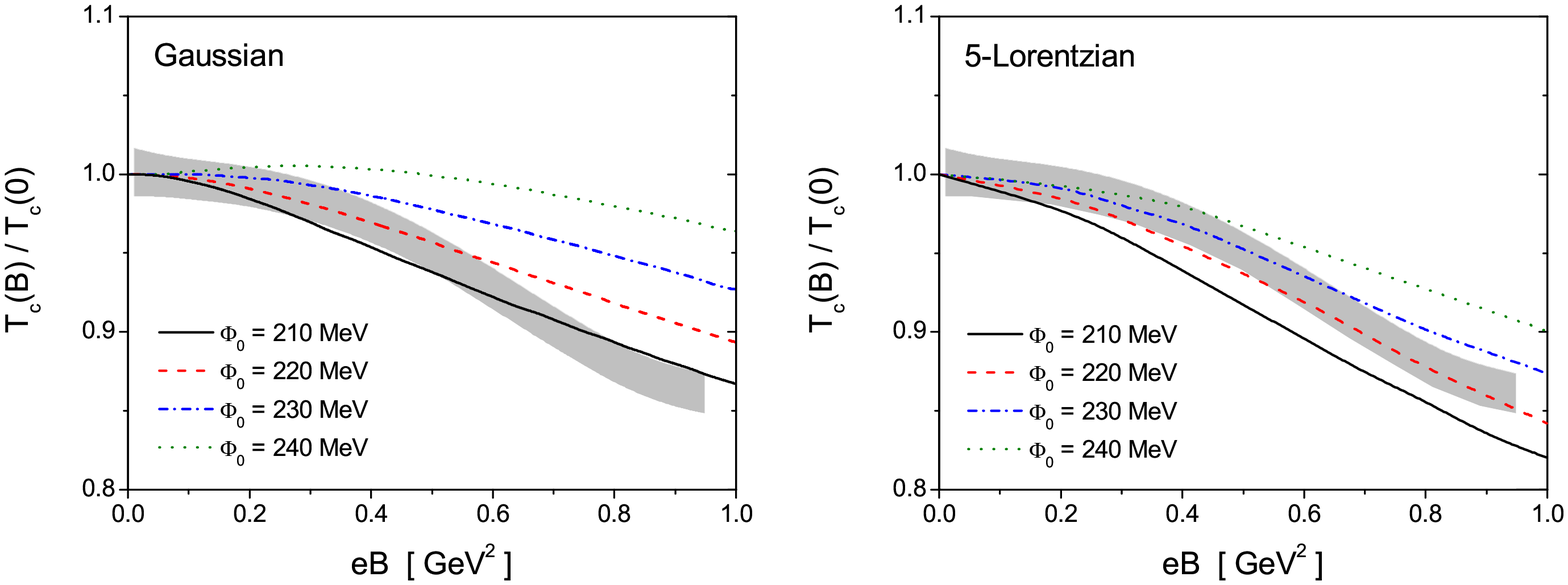}
\caption{Normalized chiral restoration temperatures as functions of $eB$ for
various model parametrizations. For comparison, LQCD results of
Ref.~\cite{Bali:2012zg} are indicated by the gray band. Left and right
panel correspond to Gaussian and 5-Lorentzian form factors, respectively.}
\label{fig3}
\end{figure}

We turn now to our numerical results for the case of finite temperature. In
the left panel of Fig.~\ref{fig2} we quote the values obtained for $\Delta
\bar \Sigma_{B,T}$ as a function of $eB$, for some representative values of
the temperature, while in the right panel we show the results for
$(\Sigma^u_{B,T}+\Sigma^d_{B,T})/2$ as a function of $T$, for some selected
values of $eB$. All these values correspond to the parametrization leading
to $\Phi_0 = 230$ MeV, yet qualitatively similar results are found for the
other parametrizations under consideration. The plots in the left panel
clearly show that, in contrast to what happens at zero temperature, the
quantity $\Delta \bar \Sigma_{B,T}$ does not display a monotonous increase
with $eB$ when one approaches the chiral transition temperature [for this
parameter set one has $T_c(eB=0) = 129.8$ MeV]. In fact, the curves reach a
maximum after which $\Delta \bar \Sigma_{B,T}$ starts to decrease with
increasing $eB$, implying that the present nonlocal model naturally exhibits
the IMC effect found in LQCD. This feature can also be seen from the results
displayed in the right panel of Fig.~\ref{fig2}. As expected, all curves
show a crossover transition from the chiral symmetry broken phase to the
(partially) restored one as the temperature increases. However, contrary to
what happens e.g.~in the standard local NJL
model~\cite{Kharzeev:2012ph,Andersen:2014xxa,Miransky:2015ava}, it is seen
that within the present model the transition temperature decreases as the
magnetic field increases. To be more specific, let us define the critical
transition temperature as the value of $T$ at which the absolute value of
the derivative $\partial[(\Sigma^u_{B,T}+\Sigma^d_{B,T})/2]/\partial T$
reaches a maximum. Since, as known from previous
analyses~\cite{Contrera:2007wu,GomezDumm:2004sr,General:2000zx}, the present
model is too simple so as to provide realistic values for the critical
temperatures even at vanishing external magnetic field, for comparison with
LQCD calculations we consider the relative quantity $T_c(B)/T_c(0)$. The
results corresponding to the previously considered parametrizations are
shown in the left panel of Fig.~\ref{fig3}, together with LQCD results from
Ref.~\cite{Bali:2012zg}. From the figure it is clearly seen that for
magnetic fields beyond $eB \simeq 0.4$~GeV$^2$ all parameter sets lead to a
decrease of the critical temperature when $eB$ gets increased, i.e.~in all
cases the IMC effect is observed. In fact, only for the case of $\Phi_0 =
240$ MeV a slightly opposite behavior is found for lower values of $eB$. On
the other hand, the strength of the IMC effect is rather sensitive to the
parametrization, the best agreement with LQCD being obtained for the
parameter set associated with the lowest value of $\Phi_0$ considered here.
Finally, in order to have some estimate of the dependence of our results on
the nonlocal form factor, in the right panel of Fig.~3 we quote the curves
corresponding to the relative transition temperatures for the case of a
5-Lorentzian function $g(p^2) = [1 + (p^2/\Lambda^2)^5]^{-1}$, often used in
the literature. It can be seen that once again the IMC effect is observed
for various parametrizations, allowing one to get a reasonably good
agreement with LQCD results.

To shed some light on the mechanism that leads to the IMC effect in our
model it is worth noticing that the nonlocal form factor turns out to be a
function of the external magnetic field. This can be clearly seen from
Eq.~(\ref{mpk}). In addition, it is important to take into account that in
nonlocal NJL-like models the form factors play the role of some finite-range
gluon-mediated effective interaction. Thus, the magnetic field dependence of
the form factor can be understood as originated by the backreaction of the
sea quarks on the gluon fields. It is interesting to consider the effective
mass for the particular case of a Gaussian form factor, given by
Eq.~(\ref{gauss}). It can be seen that in this case the components of the
momentum that are parallel and transverse to the magnetic field become
disentangled. While for the $3,4$ components the original exponential form
$\exp{(-\bar p^2/\Lambda^2)}$ is maintained, the $1,2$ (transverse) part
leads to a factor given by a ratio of polynomials in $|q_f B|/\Lambda^2$,
which goes to zero for large $B$. One might try to interpret such a factor
as a sort of effective magnetic dependent coupling constant, in the line of
the analysis carried out e.g.\ in Ref.~\cite{Ferreira:2014kpa} in the
framework of the local NJL model. However, the analogy is limited by the
fact that, contrary to what happens in the case of the local NJL, in our
model the so-defined effective coupling is not unique but depends on the
Landau level. This important difference prevents a detailed comparison with
the particular functional forms used in local NJL analyses. In any case, the
qualitative relevant feature is that for any value of $k$ the strength of
the effective coupling decreases as $eB$ increases. This is analogous to
what happens with the $B$-dependent coupling constants considered e.g.\ in
Refs.~\cite{Ferreira:2014kpa,Farias:2014eca}, and thus the IMC effect can be
understood on these grounds.

\hfill

\noindent {\em Summary and outlook}

In this work we have studied the behavior of strongly interacting matter
under an external homogeneous magnetic field in the context of nonlocal
chiral quark models. These theories are a sort of nonlocal extensions of the
local NJL model, intending to represent a step further toward a more
realistic modelling of QCD. Considering a Gaussian nonlocal form factor, we
have found that at zero temperature the behavior of the quark condensates
under the external field shows the expected magnetic catalysis effect, our
predictions being in good quantitative agreement with LQCD results. On the
other hand, in contrast to what happens in the standard local NJL model at
the mean field level, when the analysis is extended to the case of finite
temperature our results show that nonlocal models naturally lead to the
Inverse Magnetic Catalysis effect. It is worth stressing that in these
models the current-current couplings turn out to be dependent on the
temperature and the magnetic field through the nonlocal form factors, which
in principle follow from some finite-range gluon-mediated effective
interaction. Our results indicate that this scheme seems to capture the main
features of more sophisticated approaches to the QCD dynamics in the
presence of external magnetic fields, in which IMC is observed. We have also
analyzed the numerical results obtained for other form factor shapes often
considered in the literature (see e.g.\ Ref.~\cite{GomezDumm:2006vz}). For
comparison we have quoted in this work the results for the relative critical
temperatures corresponding to a 5-Lorenztian form factor, which also show
the presence of the IMC effect. A further analysis of the predictions
arising from other possible form factors, together with a more extended
presentation of the formalism, will be provided in a forthcoming
article~\cite{Pag16}. It is also worth noticing that, as a first step in
this research line, we have considered here a simple version of nonlocal
models in which e.g.~we have not incorporated interactions leading to quark
wave function renormalization nor the coupling of fermions to the Polyakov
loop. It is clear that the inclusion of these interactions is important to
provide a more realistic description of strong interaction
thermodynamics~\cite{Contrera:2007wu,Hell:2008cc}. We expect to report
progresses in this direction in the near future.

\hfill

\noindent {\em Acknowledgements}

This work has been supported in part by CONICET and ANPCyT (Argentina),
under grants PIP14-492, PIP12-449, and PICT14-03-0492, by the National
University of La Plata (Argentina), Project No.\ X718, by the Mineco (Spain)
under contract FPA2013-47443-C2-1-P, by the Centro de Excelencia Severo
Ochoa Programme grant SEV-2014-0398, and by Generalitat Valenciana (Spain),
grant PrometeoII/2014/066. DGD also acknowledges financial support from
CONICET under the PVCE programme D2392/15.

\end{document}